\documentclass[12pt]{elsart}

\usepackage{graphicx}
\usepackage{subfigure}
\usepackage{amsmath}
\usepackage{cite}
\usepackage{tikz}
\usepackage{setspace}

\begin{document}

\begin{frontmatter}

\title{Residual entropy of spin-$s$ triangular Ising antiferromagnet}
\author{M. \v{Z}ukovi\v{c}}
\ead{milan.zukovic@upjs.sk}
\address{Department of Theoretical Physics and Astrophysics, Faculty of Science,\\ 
P.J. \v{S}af\'arik University, Park Angelinum 9, 041 54 Ko\v{s}ice, Slovakia}

\begin{abstract}
We employ a thermodynamic integration method (TIM) to establish the values of the residual entropy for the geometrically frustrated spin-$s$ triangular Ising antiferromagnet, with the spin values $s=1/2,1,3/2,2$ and $5/2$. The case of $s=1/2$, for which the exact value is known, is used to assess the TIM performance. We also obtain an analytical formula for the lower bound in a general spin-$s$ model and conjecture that it should reasonably approximate the true residual entropy for sufficiently large $s$. Implications of the present results in relation to reliability of the TIM as an indirect method for calculating global thermodynamic quantities, such as the free energy and the entropy, in similar systems involving frustration and/or higher spin values by standard Monte Carlo sampling are briefly discussed.
\end{abstract}


\end{frontmatter}

\section{Introduction}
Frustrated spin systems are of considerable interest owing to their remarkable properties, such as lack of ordering at zero temperature and highly degenerate ground states (GS) with a non-vanishing entropy.~\cite{lieb,simo,moes}. However, due to frustration resulting in the high degeneracy, evaluation of thermodynamic properties in such systems is a difficult matter. For example, an exact analytical solution for the density of states is almost impossible. Then one has to resort to numerical calculations in order to calculate thermodynamic quantities such as free energy and entropy. One of the simplest such models is a triangular lattice Ising antiferromagnet (TLIA). The simplest case with the spin $s=1/2$ was treated exactly already in the 1950's~\cite{wann,hout} and has been shown to posses a large ground-state degeneracy resulting in the residual entropy density (entropy per spin) $S_0/N=0.32306$.  \\
\hspace*{5mm} Several studies~\cite{naga,yama,lipo,zeng} have shown that the ground-state properties of TLIA can be significantly affected by the spin value and can lead to long-range ordering for sufficiently large $s$. Nevertheless, for such systems there are no exact results that would characterize the ground state manifold (GSM) and, to our best knowledge, there have neither been any attempts to establish the values of the residual entropy density by numerical means. Both exact calculations
based on exhaustive scanning of the whole GSM as well as numerical techniques based on the analysis of a representative set of states belonging to GSM become practically impossible on approach to the thermodynamic limit due to exponentially growing number of states belonging to GSM. For the $\pm J$ Ising model - a typical complex system featuring a high degree of frustration and disorder - there have been various numerical approaches to determine thermodynamic properties such as entropy ~\cite{kirk,vann,morg,cheu,kola,hart}. Particularly, the thermodynamic integration method (TIM) in ref.~\cite{kirk} has been shown to give competitive results in comparison with the other methods~\cite{roma}. \\
\hspace*{5mm} In the present Letter we apply the TIM to establish the values of the residual entropy density for the geometrically frustrated spin-$s$ TLIA model for the spin values $s=1/2,1,3/2,2$ and $5/2$, based on Monte Carlo simulation results. We also formulate an analytical expression for its lower bound for general spin $s$.

\section{Model and methods} 
\hspace*{5mm} We consider the spin-$s$ Ising model on a lattice with triangular geometry described by the Hamiltonian 
\begin{equation}
\label{Hamiltonian}
\mathcal H=-J\sum_{\langle i,j \rangle}s_{i}s_{j},
\end{equation}
where the spins on the $i$th lattice site are allowed to take $2s+1$ values: $s_{i}=-s,-s+1,\ldots,s-1,s$. The summation $\langle i,j \rangle$ runs over nearest-neighbor sites and $J<0$ is an antiferromagnetic exchange interaction parameter (without loss of generality we put $J=-1$ throughout the Letter). 
\subsection{Lower bound} 
\begin{figure}[t]
\centering
\includegraphics[scale=0.65,clip]{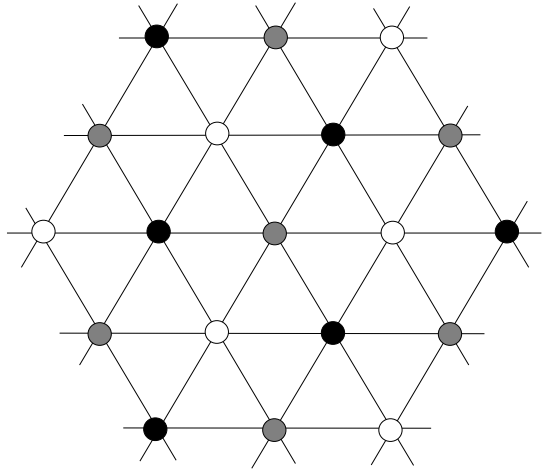}
\caption{Ground-state configuration with $1/3$ of the spins in the state $s$ (black symbols), $1/3$ in the state $-s$ (white symbols) and $1/3$ ``free'' (gray symbols).}\label{fig:schem_a}
\end{figure}

In order to roughly estimate the number of degenerate states in the infinite lattice, let us identify the states with the highest contributions. We follow Wannier's approach~\cite{wann} and apply it to the system with general spin $s$. Focusing on the elementary triangle, the lowest energy $E=-s^2$ is realized by the spin arrangement when two bonds are satisfied and the spins take the extremal values $\pm s$, such as, for example $(s_i,s_j,s_k)=(s,-s,-s)$. However, if the neighboring triangles form a hexagonal plaquette with the alternating spin values $+s$ and $-s$ around the circumference, as shown in fig.~\ref{fig:schem_a}, then the central spin is free to take independently any of $2s+1$ values without change of energy. Thus we have at least $(2s+1)^{N/3}$ degenerate states corresponding to the lowest energy. Such arrangements include cases when some neighboring free spins, which are next-nearest neighbors on the lattice and lie on the same hexagonal plaquettes, are in the same state $s$ or $-s$ alternating around the the circumference with the spins in the states $-s$ or $s$, respectively. Such plaquettes produce additional $\frac{2N}{3(2s+1)^3}$ free spins and thus the total number of degenerate states increases to $(2s+1)^{\frac{N}{3}\big[1+\frac{2}{(2s+1)^3}\big]}$. Of course, there may be further more involved contributions, but limiting ourself to the above considerations we can estimate the lower bound on the residual entropy density given by (putting $k_B=1$)
\begin{equation}
\label{entropy_lb}
S_0/N \ge \frac{1}{3}\Big[1+\frac{2}{(2s+1)^3}\Big]\ln(2s+1).
\end{equation}

\subsection{Thermodynamic integration method (TIM)} 
\begin{figure*}[t]
\centering
\subfigure[]{\includegraphics[scale=0.55]{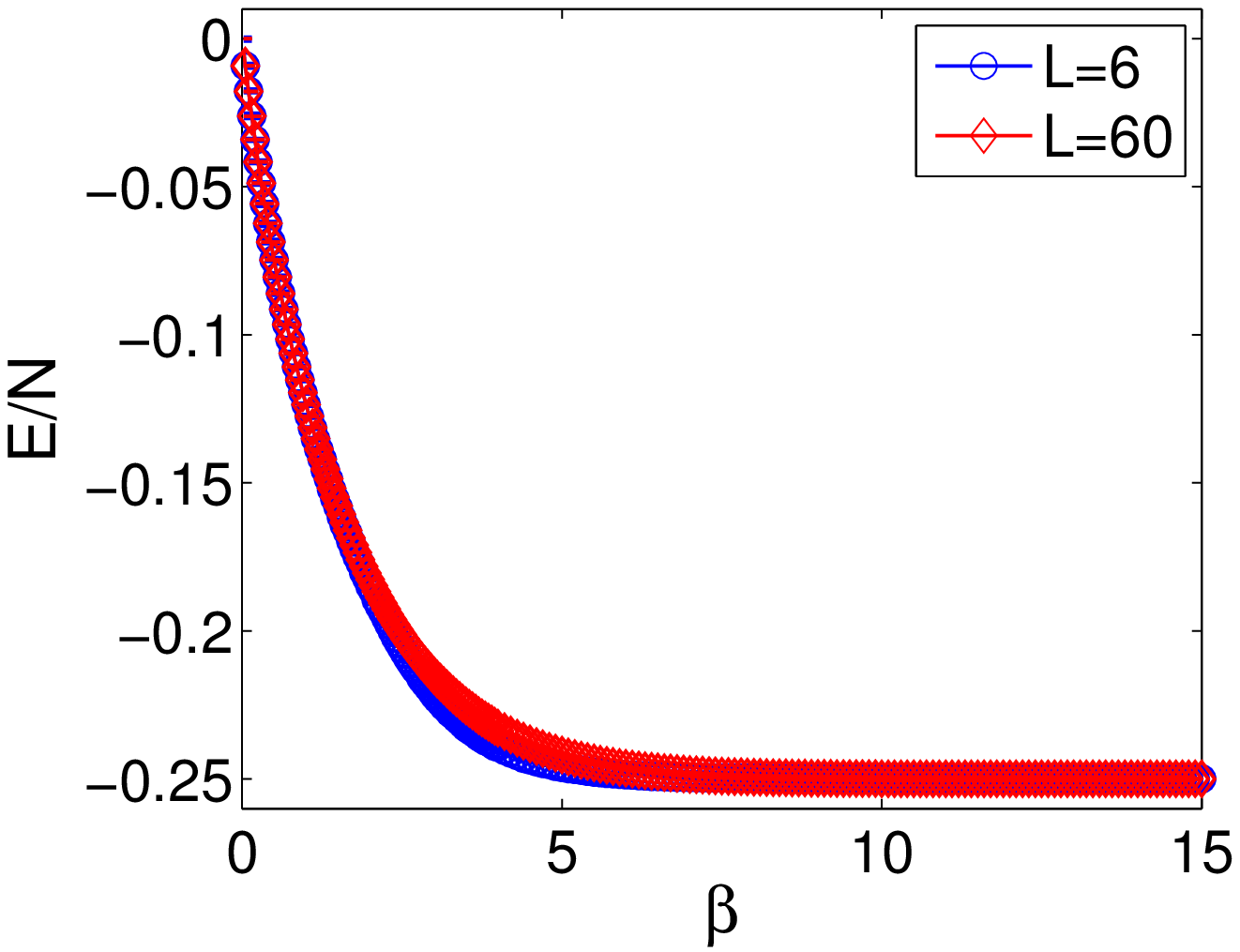}\label{fig:spin1_2a}}
\subfigure[]{\includegraphics[scale=0.55]{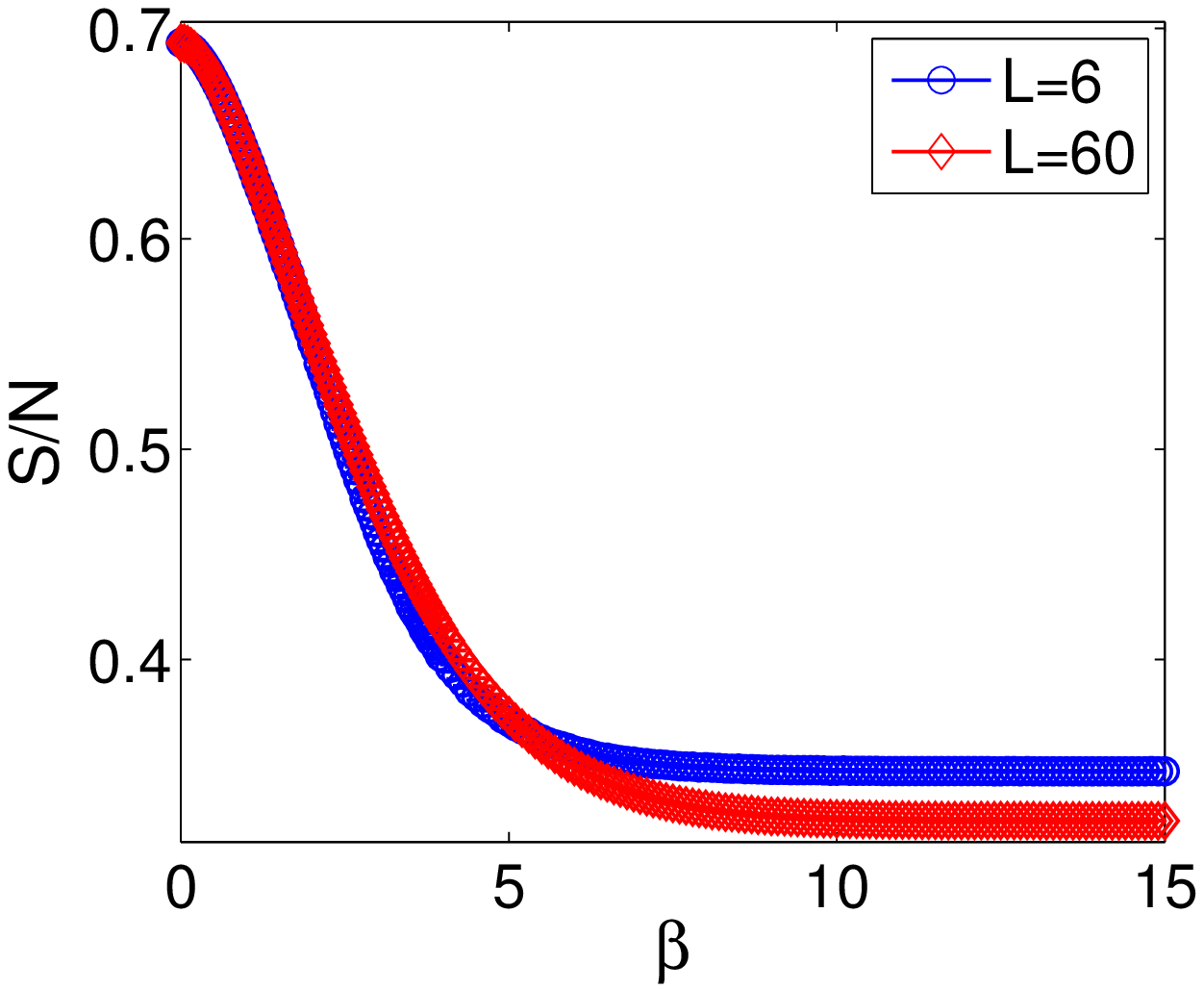}\label{fig:spin1_2b}}
\subfigure[]{\includegraphics[scale=0.55]{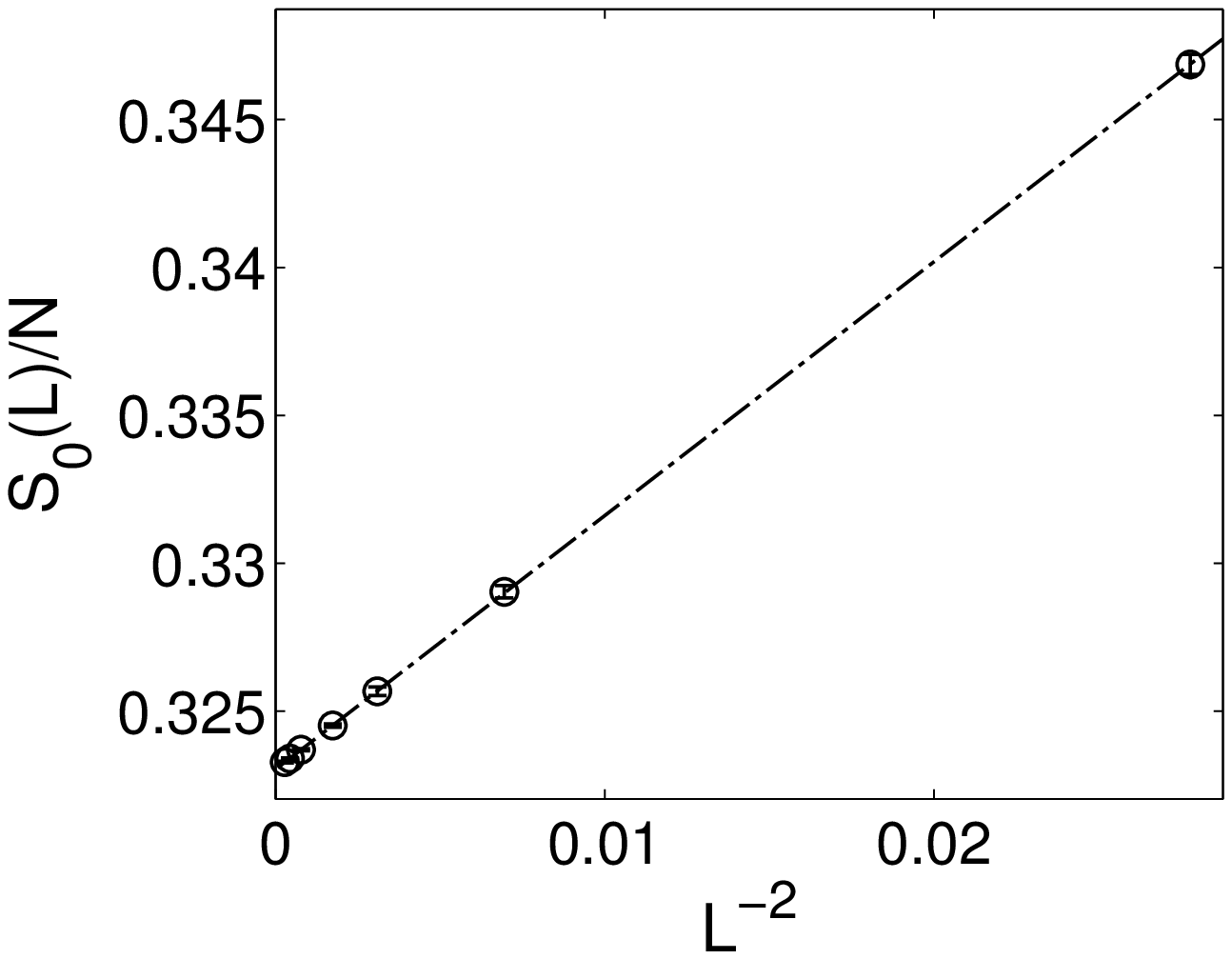}\label{fig:spin1_2c}}
\caption{(Colour on-line) Inverse temperature dependence of (a) the internal energy density, (b) entropy density and (c) FSS of the residual entropy density for spin $s=1/2$. In all figures the error bars are smaller than the symbol sizes.}\label{fig:spin1_2}
\end{figure*}

The TIM is based on Monte Carlo (MC) simulation results for either the internal energy or the specific heat, as described below.
We perform MC simulations on a spin system of linear size $L$, employing the Metropolis dynamics and applying the periodic boundary conditions to eliminate boundary effects. For thermal averaging we consider $N_{MC}$ Monte Carlo sweeps (MCS) or steps per spin after discarding another $N_0$ (we take $20\%$ of $N_{MC}$) MCS for thermalization. The simulations start from zero inverse temperature $\beta\equiv 1/T$ ({\it i.e}., $T = \infty$), using a random initial configuration, and proceed up to some $\beta_{max}$. $\beta$ is increased by the step $\Delta \beta$ and the simulation at $\beta+\Delta \beta$ starts from the final configuration obtained at $\beta$. We calculate the internal energy $E=\langle H \rangle$, which is used in the TIM for estimation of the residual entropy~\cite{kirk,bind}. Typically, the TIM is based on integrating the expression $dS(T)=C/TdT$, where $S(T)$ is the entropy at temperature $T$ and $C$ is the specific heat. Using the knowledge of the entropy in the infinite temperature limit\footnote{In the spin-$s$ Ising model all the $(2s+1)^N$ possible configurations are equally likely and thus $S(T=\infty)=N\ln (2s+1)$.}, we obtain the formula for $S(T)$ by integrating from infinity: 
\begin{equation}
\label{TIM_C}
S(T)= N\ln (2s+1) - \int_{\infty}^{T}{\frac{C(T')}{T'}}dT'.
\end{equation}
An alternative approach, which generally reduces the errors resulting from the calculation of the specific heat $C$ from the energy $E$ and its numerical integration, particularly in case it shows a sharp peak near the transition temperature, is based on integration of the internal energy, a quantity which is directly measured in MC simulations. The formula equivalent to eq.~(\ref{TIM_C}) is given by
\begin{equation}
\label{TIM_E}
S(\beta)= N\ln (2s+1) + \beta E(\beta)- \int_{0}^{\beta}E(\beta')d(\beta').
\end{equation}
Then the residual entropy $S_0$ is estimated by taking $\beta$ to infinity. Then also the free energy can be obtained as $F(\beta)=E(\beta)-S(\beta)/\beta$. \\
\hspace*{5mm} In practice, however, the numerical quadrature on r.h.s of eq.~(\ref{TIM_E}) instead of infinity is performed to some value $\beta_{max}$ beyond which the system is expected to remain in the GSM during the observation time. Then, upon further increase of $\beta$ the internal energy shows virtually no fluctuations and thus the entropy practically does not change (see fig.~\ref{fig:spin1_2}). In our tests we determined $\beta_{max}=15$ (corresponding to $T \approx 0.067$). Since the quadrature error is inversely proportional to the square of the number of measurements~\cite{roma}, we set the step $\Delta \beta$ sufficiently small (smaller in the region where the energy variation is larger) to achieve enough measurements and the resulting quadrature error of order of $10^{-5}$. 
The number of measurements along with the thermalization and simulation times are the parameters that affect the simulation error. Finally, for each lattice size it is desirable to perform $M$ simulation runs to suppress the sampling error. Targeting reasonably small final errors but also considering the available computational resources, we chose the following parameter values: $N_{MC}=5 \times 10^4$, $M=20$ for smaller and $N_{MC}= 10^5$, $M=10$ for larger lattices. To eliminate finite-size effects, the simulations are performed on lattices with $L=6$ up to $L=60$ and the thermodynamic limit value is estimated by extrapolation $L \rightarrow \infty$ in a finite-size scaling (FSS) analysis.

\section{Results}
\begin{figure*}[t]
\centering
\subfigure[$s=1$]{\includegraphics[scale=0.55]{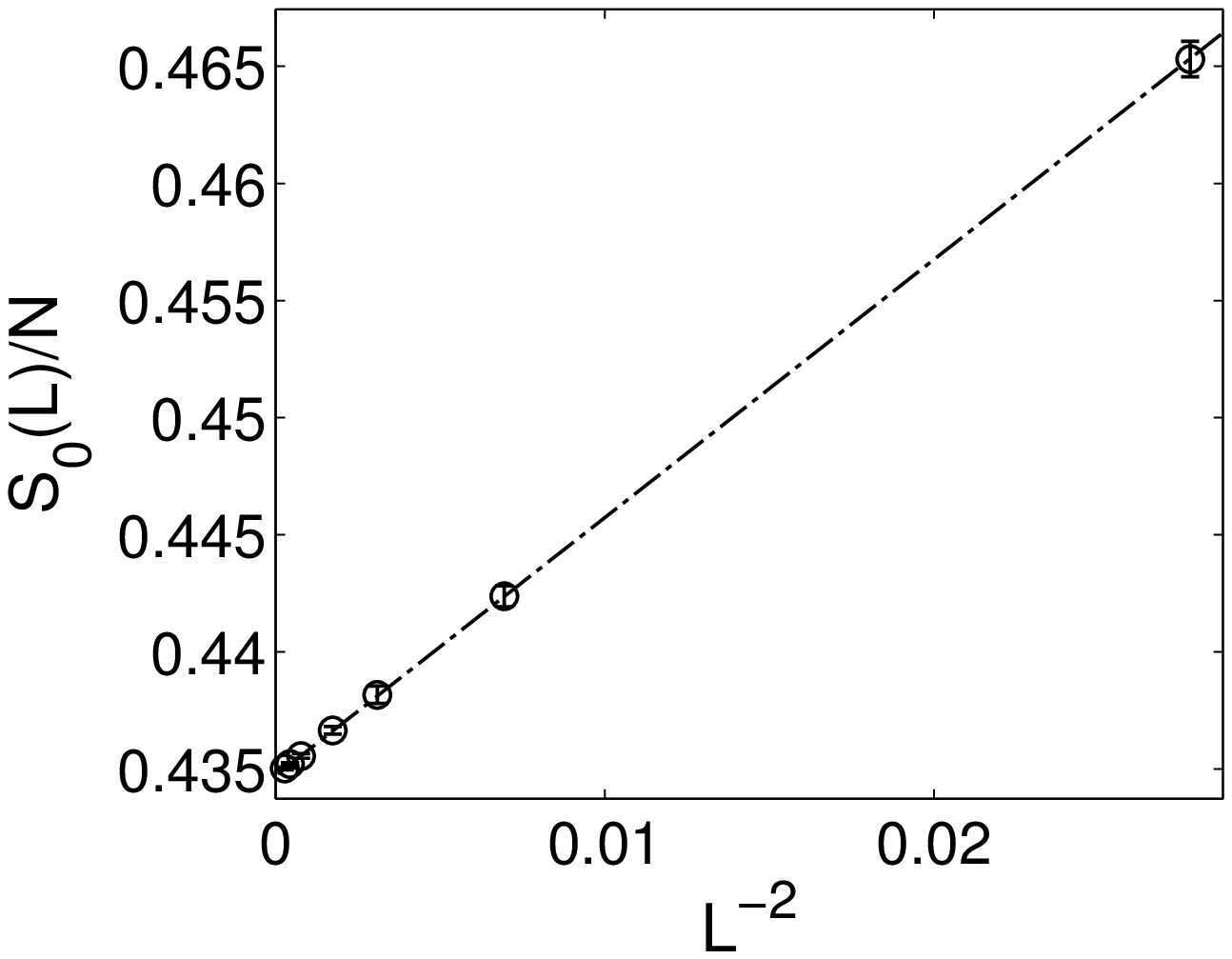}}
\subfigure[$s=3/2$]{\includegraphics[scale=0.55]{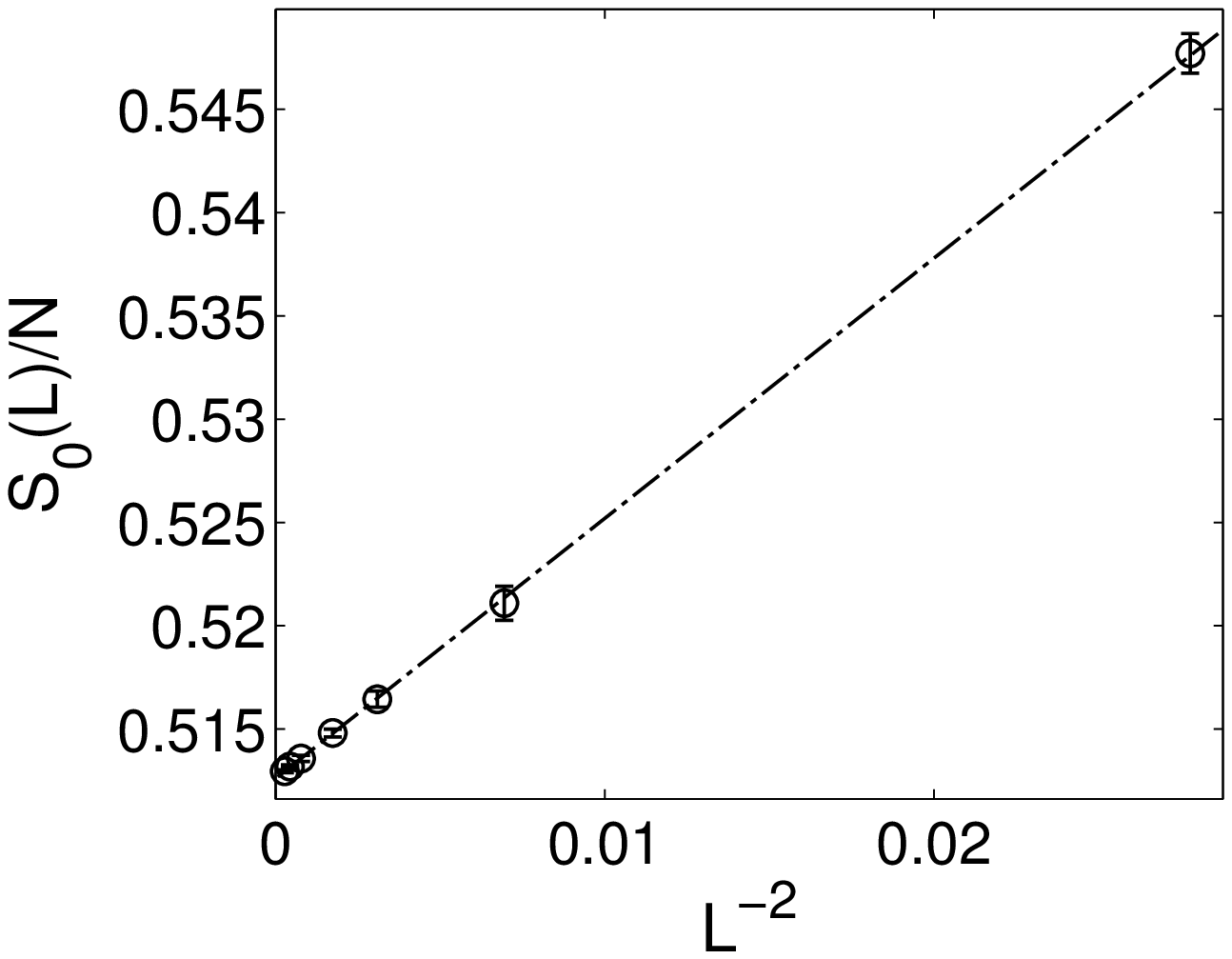}}
\subfigure[$s=2$]{\includegraphics[scale=0.55]{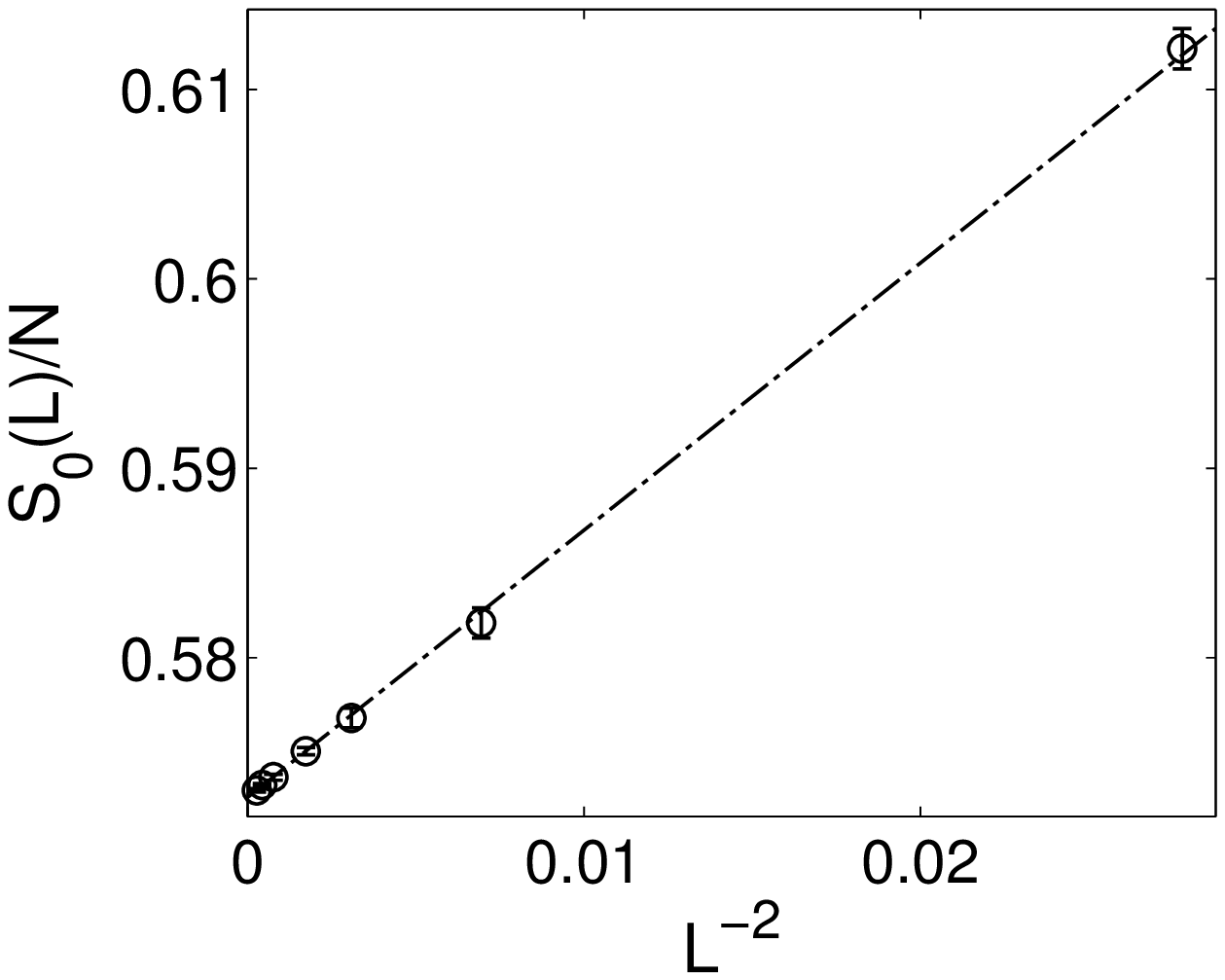}}
\subfigure[$s=5/2$]{\includegraphics[scale=0.55]{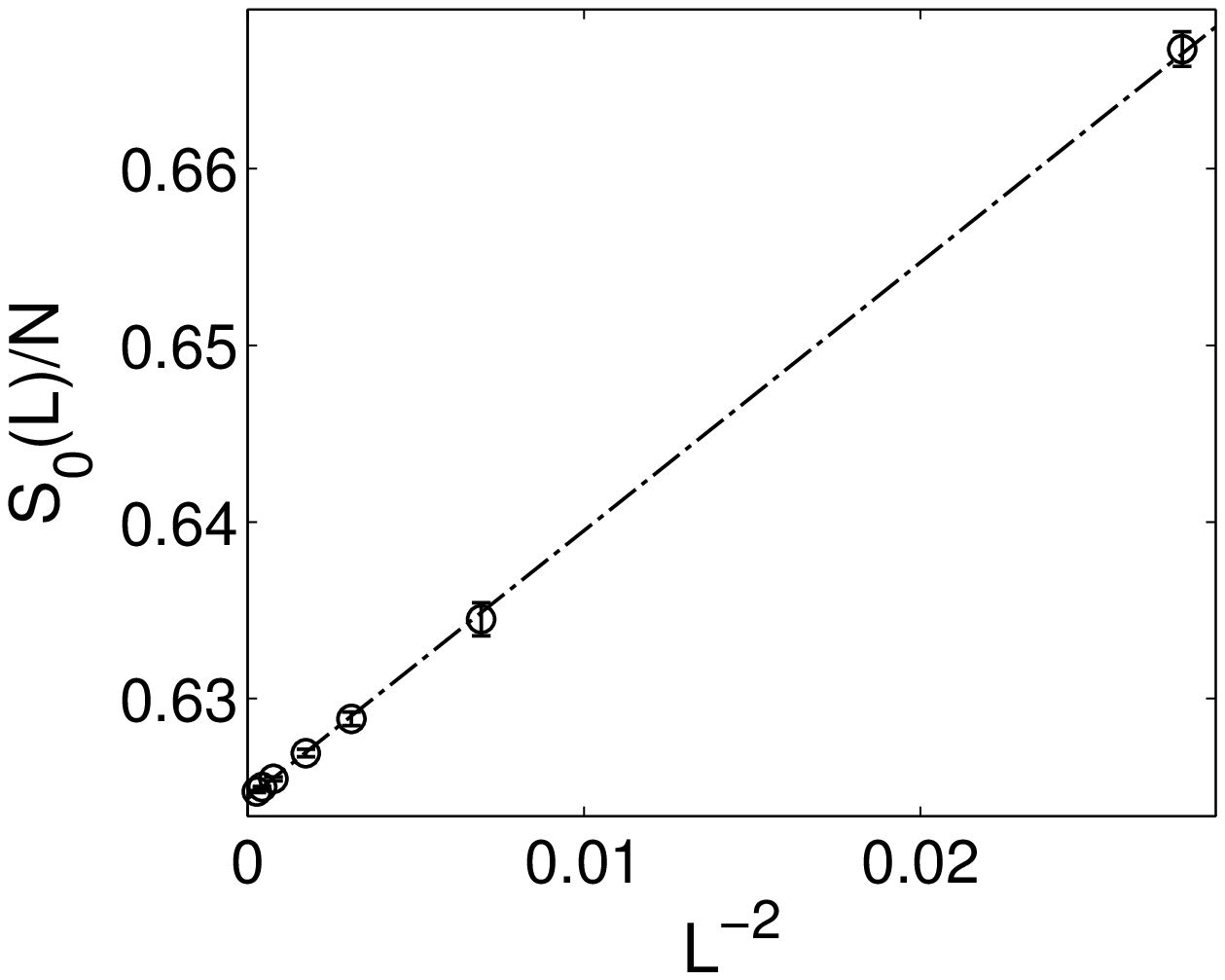}}
\caption{FSS of the residual entropy density for the spin values $s=1,3/2,2$ and $5/2$.}\label{fig:all}
\end{figure*}

\begin{figure}[]
\centering
\includegraphics[scale=0.65]{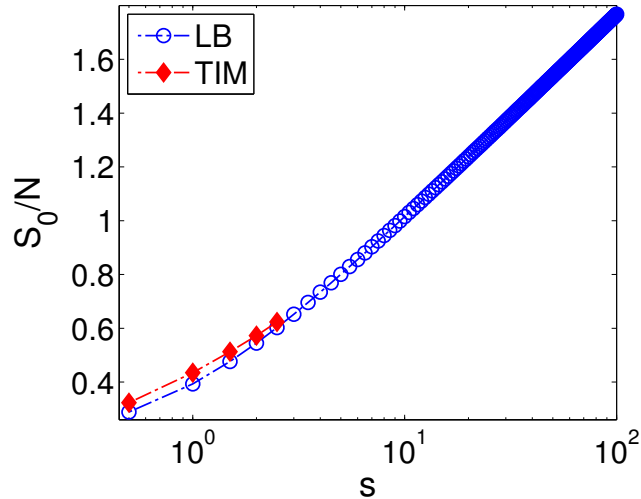}
\caption{(Colour on-line) Variation of the residual entropy density with the spin value. The empty circles represent the lower bounds and the filled diamonds the values obtained from TIM. The error bars of $S_0^{TIM}/N$ values are smaller than the symbol sizes.}\label{fig:re_plot}
\end{figure}

Below we present the results for the following spin values $s=1/2,1,3/2,2$ and $5/2$. For $s=1/2$, the residual entropy density $S_0/N = 0.32306$~\cite{wann,wann2} is known exactly (we note that the value in ref.~\cite{wann} is incorrect), however, we include it in our calculations to verify the reliability of the TIM. First of all, from relation (\ref{entropy_lb}) let us estimate the lower bounds of the residual entropy densities for the respective spin values (see table~\ref{tab:re_ent}). The values increase with the increasing $s$ (for larger $s$ the increase is virtually logarithmic) and it is easy to see that they tend to infinity for $s \to \infty$. 
\begin{table}
\caption{Residual entropy densities $S_0^{TIM}/N$ estimated by the TIM and their lower bounds $S_0^{LB}/N$ obtained from equation~(\ref{entropy_lb}), for the spin values $s=1/2,1,3/2,2$ and $5/2$. $\bar{R}^2$ represents an adjusted coefficient of determination~\cite{theil} of the linear fit $S_0(L)/N$ vs $L^{-2}$.}
\label{tab:re_ent}
\begin{center}
\begin{tabular}{cccccccc}
$s$            & 1/2  & 1 & 3/2 & 2 & 5/2 & $\ldots$ & $\infty$\\
\hline
$S_0^{LB}/N$    &   0.28881 & 0.39333  & 0.47654  &  0.54506  & 0.60278 & $\ldots$ & $\infty$ \\
$S_0^{TIM}/N$            &   0.32303(2) & 0.43472(4)  & 0.51262(5)  &  0.57263(6)  & 0.62331(6) & $\ldots$ & $-$ \\
$\bar{R}^2$            &   1.0000 & 1.0000  & 0.9999  &  0.9995  & 0.9998 & $\ldots$ & $-$ \\
$\frac{(S_0^{TIM} - S_0^{LB})}{S_0^{TIM}} 100 \%$  &  10.6  &  9.5  & 7.0  & 4.8  & 3.3  &  $\ldots$ & $-$ \\
\end{tabular}
\end{center}
\end{table}
\\
\hspace*{5mm} Next, let us estimate the residual entropy densities for the respective spin values by applying the TIM. In fig.~\ref{fig:spin1_2} we present the results for the case of $s=1/2$. Fig.~\ref{fig:spin1_2a} shows the inverse temperature $\beta$ dependence of the internal energy density for the smallest ($L=6$) and the largest ($L=60$) lattice sizes. The corresponding values of the entropy densities are evaluated using expression~(\ref{TIM_E}) and presented in fig.~\ref{fig:spin1_2b}. The error bars are smaller than the symbol sizes. For large values of $\beta$ (low enough temperatures) the entropy density displays virtually no change and thus we approximate the value of the residual entropy density as $S_0(L)/N \approx S(\beta=15,L)/N$ for each $L$. Finally, extrapolation to the thermodynamic limit is presented in fig.~\ref{fig:spin1_2c}. We obtain an excellent linear fit for entire range of the considered lattice sizes, with the adjusted coefficient of determination~\cite{theil} $\bar{R}^2=1.0000$. The estimated value of the residual entropy density $S_0^{TIM}/N=0.32303(2)$ is very close to the Wanier's exact value $0.32306$.
\begin{figure}[]
\centering
\includegraphics[scale=0.65,clip]{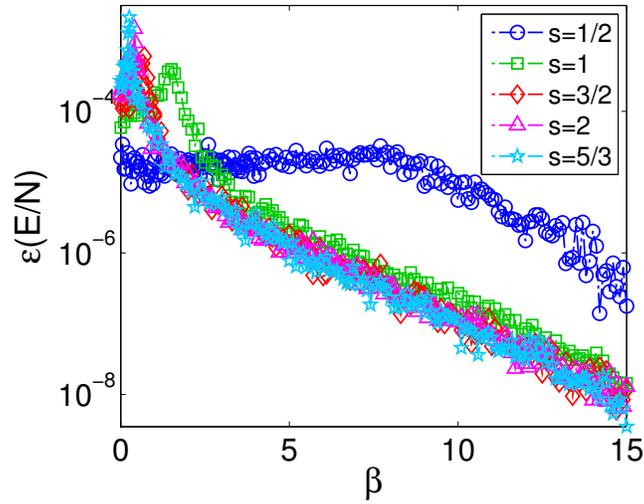}
\caption{(Colour on-line) Inverse temperature dependence of the internal energy density errors obtained from $M$ independent MC runs for $s=1/2,1,3/2,2$ and $5/2$, with $L=60$.}\label{fig:en_err}
\end{figure}
\\
\hspace*{5mm} We perform similar analysis also for the spin values $s=1,3/2,2$ and $5/2$. In MC simulations of spin systems with greater spin values one should be more careful about thermalization times and statistical errors. Owing to the adopted approach that the simulation at $\beta+\Delta \beta$ starts from the final configuration obtained at $\beta$ and taking $\Delta \beta$ sufficiently small, once the system is equilibrated at high temperatures it is maintained close to the equilibrium along the entire temperature path. Thus no long thermalization times are required and the ones we chose turned out to be more than generous. The internal energy density errors, $\epsilon(E/N)$, obtained as standard deviations from $M$ independent MC runs, were found for all the spin values too small to be visible on the internal energy scale and, therefore, we plot them separately in fig.~\ref{fig:en_err} for the largest lattice size $L=60$. Comparing the $S>1/2$ cases with the $S=1/2$ one, we can see two regimes. At higher temperatures the fluctuations are larger in the former case and show an anomalous increase around some $\beta$ but still within acceptable limits. On the other hand, at lower temperatures they are of about two orders smaller than for the $S=1/2$ case. The FSS analysis for the values $s>1/2$ are presented in fig.~\ref{fig:all}. Despite slightly larger errors than for $s=1/2$, we again obtain excellent linear scaling of $S_0(L)/N$ vs $L^{-2}$ with the adjusted coefficients of determination $\bar{R}$ close to one. These are listed in table~\ref{tab:re_ent} along with the estimated values of $S_0/N$. By comparing the values of $S_0/N$ obtained from TIM with the predictions of their lower bounds we can observe that their relative differences diminish with the increasing spin value and vanish in infinity. Thus, assuming that $S_0^{TIM}/N$ are good estimators of the true values, for the systems with spin values larger than $5/2$ the estimates of the residual entropy densities from equation~(\ref{entropy_lb}) are expected to deviate by less than $3.3 \%$ from the exact value. Variations of the residual entropy densities obtained from TIM and their lower bounds with the spin value are ploted in fig.~\ref{fig:re_plot}.   

\section{Conclusions}
Monte Carlo method was used to calculate internal energies of the geometrically frustrated spin-$s$ triangular lattice Ising antiferromagnet (TLIA) for the spin values $s=1/2,1,3/2,2$ and $5/2$. Subsequently, these were used in the thermodynamic integration method (TIM) to establish the values of the respective residual entropy densities for various lattice sizes and extrapolated to the thermodynamic limit. The result for an exactly solvable case of $s=1/2$ demonstrated a good performance of the TIM in predicting the residual entropy value. Furthermore, we obtained an analytical expression for the lower bound in a general spin-$s$ TLIA model, which is expected to reasonably approximate the residual entropy densities for larger values of $s$. \\
\hspace*{5mm} Finally, global quantities, such as the entropy or the free energy, are useful in different kinds of investigations. For example, the entropy can be used to study the magnetocaloric effect, which is particularly enhanced in frustrated systems~\cite{zito}, and its residual value can serve as a measure of frustration in the system. On the other hand, the free energy can, for example, help to determine phase transitions, particularly the first-order ones~\cite{bind1}. Nevertheless, these quantities cannot be obtained directly from standard MC simulations techniques and usually one has to retreat to some variation of algorithmically more demanding non-Boltzmann, such as Wang-Landau~\cite{wang}, MC technique. Our results indicate that this may not be necessary even when we deal with frustrated and larger-spin systems, since the indirect calculation by the TIM can give satisfactory results using conventional MC techniques\footnote{For systems with larger spin values a single spin flip algorithm, such as Metropolis, may not be the best choice and hybridizing it with cluster flipping could substantially reduce the statistical errors and lead to a further increase in performance ~\cite{plas}.} at moderate computational cost. In particular, it is convenient to use eq.~(\ref{TIM_E}) instead of~(\ref{TIM_C}). Then, the only quantity we need is the directly measured internal energy, which is generally better behaved that the specific heat for the purpose of numerical integration and, unlike some other thermodynamic functions, due to high degeneracy shows relatively small fluctuations even in frustrated systems. We believe that the present results will stimulate further similar studies on some other frustrated and/or large-spin systems. 

%
%
%
%
%

\section*{Acknowledgments}
This work was supported by the Scientific Grant Agency of Ministry of Education of Slovak Republic (Grant No. 1/0234/12). The authors acknowledge the financial support by the ERDF EU (European Union European regional development fund) grant provided under the contract No. ITMS26220120005 (activity 3.2).


\begin{thebibliography}{30}
\bibitem{lieb} R. Liebmann, Statistical Mechanics of Periodic Frustrated Ising Systems (Springer-Verlag, Berlin, 1986).
\bibitem{simo} B. Simon, The Statistical Mechanics of Lattice Gases (Princeton University Press, Princeton, NJ, 1993), Vol. 1.
\bibitem{moes} R. Moessner and S. L. Sondhi, Phys. Rev. B 63 (2001) 224401.
\bibitem{wann} G.H. Wannier, Phys. Rev. 79 (1950) 357.
\bibitem{hout} R.M.F. Houtappel, Physica (Amsterdam) 16 (1950) 425.
\bibitem{naga} O. Nagai, S. Miyashita and T. Horiguchi, Phys. Rev. B 47 (1993) 202.
\bibitem{yama} Y. Yamada, S. Miyashita, T. Horiguchi, M. Kang and O. Nagai, J. Magn. Magn. Mater. 140-144 (1995) 1749.
\bibitem{lipo} A. Lipowski, T. Horiguchi and D. Lipowska, Phys. Rev. Lett. 74 (1995) 3888.
\bibitem{zeng} C. Zeng and C.L.Henley, Phys. Rev. B 55 (1997) 14935.
\bibitem{kirk} S. Kirkpatrick, Phys. Rev. B 16 (1977) 4630.
\bibitem{vann} J. Vannimenus and G. Toulouse, J. Phys. C 10 (1977) L537.
\bibitem{morg} I. Morgenstern and K. Binder, Phys. Rev. B 22 (1980) 288.
\bibitem{cheu} H.-F. Cheung and W. L. McMillan, J. Phys. C 16 (1983) 7027.
\bibitem{kola} A.J. Kolan and R. G. Palmer, J. Appl. Phys. 53 (1982) 2198.
\bibitem{hart} A.K. Hartmann, Phys. Rev. E 63 (2001)016106.
\bibitem{roma} F. Rom\'a, F. Nieto, E.E. Vogel and A.J. Ramirez-Pastor, J. Stat. Phys. 114 (2004) 1325.


\bibitem{bind} K. Binder, The Monte Carlo method for the study of phase transitions: A review of some recent progress, J. Comput. Phys. 59 (1985) 1.
\bibitem{wann2} G.H. Wannier, Phys. Rev. B 7 (1973) 5017.

\bibitem{theil} H. Theil, Economic Forecasts and Policy, Vol. XV of Contributions to Economic Analysis, North-Holland Pub. Co., Amsterdam, 1961.


\bibitem{zito} M.E. Zhitomirsky, Phys. Rev. B 67 (2003) 104421. 

\bibitem{bind1} K. Binder, Rep. Prog. Phys. 50 (1987) 783.
    
\bibitem{wang} F. Wang, D.P. Landau, Phys. Rev. Lett. 86 (2001) 2050.

\bibitem{plas} J.A. Plascak, A.M. Ferrenberg, D.P. Landau, Phys. Rev. E 65 (2002) 066702.
  
\end{thebibliography}
\end{document}